\documentclass[ALICE,manyauthors]{cernphprep}
\usepackage{epstopdf}
\usepackage{graphicx}
\usepackage{amssymb}
\usepackage{hyperref}
\begin{document}%
%
%
\begin{titlepage}
\PHnumber{2012-160}                 
\PHdate{12 June 2012}              
%
%
\title{K$^0_s$K$^0_s$ correlations 
in pp collisions at $\sqrt{s}=7$ TeV\\ from the LHC ALICE experiment}
\ShortTitle{K$^0_s$K$^0_s$ correlations in ALICE}   
%
\Collaboration{ALICE Collaboration%
         \thanks{See Appendix~\ref{app:collab} for the list of collaboration
                      members}}
\ShortAuthor{ALICE Collaboration}      
\begin{abstract}
Identical neutral kaon pair correlations are measured in $\sqrt{s}=7$
TeV pp  collisions in the ALICE experiment. One--dimensional
K$^0_s$K$^0_s$ correlation functions in terms of the invariant
momentum difference of kaon pairs are formed in two multiplicity
and two transverse momentum ranges. The femtoscopic 
parameters for the radius and correlation strength of the kaon source are 
extracted. The f${\rm i}$t includes quantum
statistics and f${\rm i}$nal--state interactions of the a$_0$/f$_0$
resonance. K$^0_s$K$^0_s$ correlations show an increase in radius
for increasing multiplicity and a slight decrease in radius for
increasing transverse mass, $m_{\rm T}$, as seen in $\pi\pi$ correlations in the pp system
and in heavy--ion collisions. Transverse mass scaling is observed between
the K$^0_s$K$^0_s$ and $\pi\pi$ radii. Also, the f${\rm i}$rst observation is made of the decay of 
the f$_2'$(1525) meson
into the K$^0_s$K$^0_s$ channel in pp collisions.
\end{abstract}
\end{titlepage}
%
%
\setcounter{page}{2}
\section{Introduction}

In this paper we present  results from a K$^0_s$K$^0_s$ femtoscopy
study by the ALICE experiment \cite{Aamodt:2008zz,Aamodt:2010aa} in pp
collisions at $\sqrt{s}=7$ TeV from the CERN LHC.  Identical boson femtoscopy, especially
identical charged $\pi\pi$ femtoscopy, has been used extensively over
the years to study experimentally the space--time geometry of the
collision region in high--energy particle and heavy--ion collisions
\cite{Lisa:2005a}.
Recently, the ALICE collaboration has carried out a charged $\pi\pi$
femtoscopic study for pp collisions at $\sqrt{s}=7$ TeV.
This study shows a transverse 
momentum dependence of the source radius developing with increasing 
particle multiplicity similar to the one 
observed in heavy-ion collisions \cite{Aamodt:2011kd}.
The main motivations to carry out the present K$^0_s$K$^0_s$
femtoscopic study to complement this $\pi\pi$
study are 1) to extend the transverse pair momentum range of the
charged $\pi\pi$ studies which typically cuts off at about 0.8 GeV/$c$
due to reaching the limit of particle identif${\rm i}$cation, whereas
K$^0_s$'s can easily be identif${\rm i}$ed to 2 GeV/$c$ and beyond,  2) since K$^0_s$ is
uncharged, K$^0_s$K$^0_s$ pairs close in phase space are not
suppressed by a f${\rm i}$nal-state Coulomb repulsion as is the case of
charged $\pi\pi$ pairs, 3) K$^0_s$K$^0_s$ pairs close in phase space
are additionally enhanced  by the strong f${\rm i}$nal-state interaction due to the
a$_0$/f$_0$ resonance giving a more pronounced signal,  and 4) one can,
in principle, obtain complementary information about the collision
interaction region by using different types of mesons.  
The physics advantage of items 1) and 4) is to
study transverse mass scaling, 
of the source size which is considered
a signature of collective behaviour in heavy-ion collisions \cite{Lisa:2005a}.
If $m_{\rm T}$ scaling is present, the source sizes from different species
should fall on the same curve vs. $m_{\rm T}$. Thus, comparing 
results from $\pi\pi$ and K$^0_s$K$^0_s$, particularly at very
different $m_{\rm T}$, would be a good test of this scaling.
Item 3) can be used as an advantage since the f${\rm i}$nal-state interaction of
K$^0_s$K$^0_s$ via the a$_0$/f$_0$ resonance can be
calculated with a reasonable degree of precision.
Previous
K$^0_s$K$^0_s$ studies have been carried out in LEP e$^+$e$^-$
collisions \cite{Abreu:1996hu, Schael:2004qn, lep3}, HERA  ep
collisions \cite{Chekanov:2007ev}, and RHIC Au--Au collisions
\cite{Abelev:2006gu}. Due to statistics limitations, a single set of
femtoscopic source parameters, i.e. radius, $R$, and 
correlation strength, $\lambda$, was extracted
in each of these studies. The present study is the
f${\rm i}$rst femtoscopic K$^0_s$K$^0_s$ study to be carried out a) in pp
collisions and b) in more than one multiplicity and transverse pair momentum,
$k_{\rm T}$,  range, where $k_{\rm T}=\left |\vec{p}_{T1}+\vec{p}_{T2}\right |/2$
and $\vec{p}_{T1}$ and $\vec{p}_{T2}$ are the two transverse momenta
of the K$^0_s$K$^0_s$ pair.

\section{Description of Experiment and Data Selection}
The data analyzed for this work were taken by the ALICE experiment during the
2010 $\sqrt{s}=7$ TeV pp run at the CERN LHC. About $3\times10^8$
minimum bias events were analyzed.

K$^0_s$ identif${\rm i}$cation and momentum determination were performed
with particle tracking in the ALICE Time Projection Chamber (TPC) and
ALICE Inner Tracking System (ITS) \cite{Aamodt:2008zz,Aamodt:2010aa}. 
The TPC was used to record charged-particle tracks as they left ionization 
trails in the Ne--CO$_2$ gas. The ionization drifts up to 2.5 m from the central 
electrode to the end caps to be measured on 159 padrows, which are grouped 
into 18 sectors; the position at which the track crossed the padrow was 
determined with resolutions of 2 mm and 3 mm in the drift and transverse 
directions, respectively. The ITS was used also for tracking. 
It consists of six silicon layers, two innermost Silicon Pixel Detector (SPD) layers, 
two Silicon Drift Detector (SDD) layers, and two outer Silicon Strip Detector (SSD) 
layers, which provide up to six space points for each track. The tracks 
used in this analysis were reconstructed using the information from both the 
TPC and the ITS; such tracks were also used to reconstruct the primary vertex 
of the collision. For details of this
procedure and its eff${\rm i}$ciency see Ref.\cite{Aamodt:2010aa}. 

The forward scintillator detectors, VZERO, are placed along the beam 
line at $+3$ m and $-0.9$ m from the nominal interaction point. They cover a 
region $2.8<\eta<5.1$ and $-3.7<\eta < -1.7$, respectively. They were 
used in the minimum--bias trigger and their timing signal was used to reject 
the beam--gas and beam--halo collisions.
The minimum--bias trigger required a signal in either of the two VZERO 
counters or one of the two inner layers of the SPD. Within this sample, events were selected 
based on the measured charged--particle multiplicity within the pseudorapidity 
range $\left | \eta \right |<1.2$. Events were required to have a primary 
vertex within 1 mm of the beam line and 10 cm of the centre of the 5 m long TPC. 
This provides almost uniform acceptance for particles with $\left | \eta \right |<1$ 
for all events in the sample. It decreases for $1.0<\left | \eta \right |<1.2$. 
In addition, we require events to have at least one charged particle 
reconstructed within $\left | \eta \right |<1.2$.

\begin{figure}
\begin{center}
\includegraphics[width=100mm]{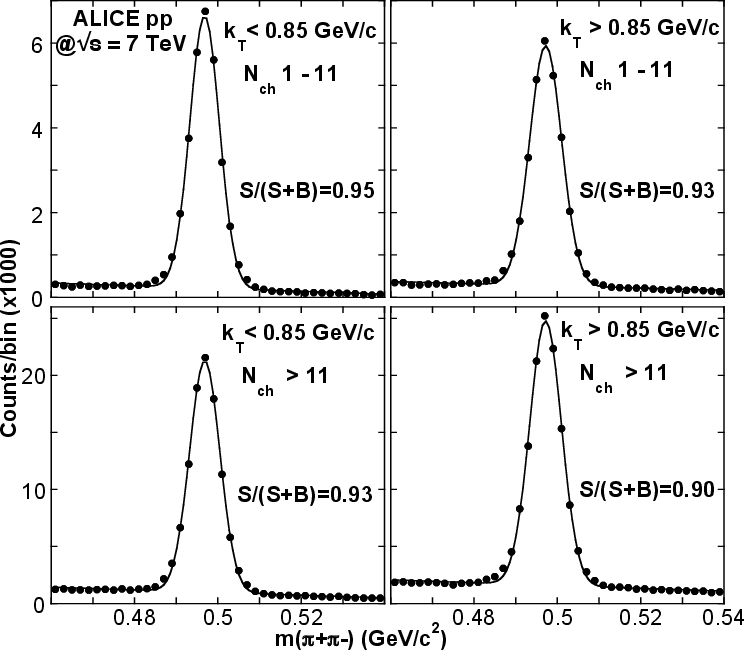} \caption{Invariant mass distributions 
of $\pi^+\pi^-$ pairs in the four multiplicity--$k_{\rm T}$ ranges used in the study. K$^0_s$ used in this 
analysis were identif${\rm i}$ed by the cut 0.490 GeV/$c^2$ $< m <$ 0.504 GeV/$c^2$.
Also shown is a Gaussian+linear f${\rm i}$t to the data points.}
\label{fig1}
\end{center}
\end{figure}

The decay
channel K$^0_s\rightarrow\pi^+\pi^-$ was used for particle
identif${\rm i}$cation, with a typical momentum resolution of $\sim1$\%
\cite{Aamodt:2011zz}.
The distance of closest approach (DCA) of the candidate K$^0_s$ decay
daughters was required to be $\leq 0.1$ cm.
Figure \ref{fig1} shows invariant mass distributions
of candidate K$^0_s$ vertices  for the four multiplicity--$k_{\rm T}$ ranges used in this 
study (see below)
along with a Gaussian + linear f${\rm i}$t to the data. The invariant mass
at the peaks was found to be 0.497 GeV/$c^2$, which is within 1
MeV/$c^2$ of the accepted mass of the K$^0_s$ \cite{pdg}. The average
peak width was $\sigma=3.72$ MeV/$c^2$ demonstrating the good K$^0_s$
momentum resolution obtained in the ALICE tracking detectors.
A vertex was
identif${\rm i}$ed with a K$^0_s$ if the invariant mass of the candidate
$\pi^+\pi^-$ pair associated with it fell in the range 0.490-0.504
GeV/$c^2$. As seen in Figure \ref{fig1}, the ratio of the K$^0_s$ 
signal to signal+background, $S/(S+B)$, in 
each of the four ranges
is determined to be 0.90 or greater. The minimum K$^0_s$ f${\rm l}$ight distance
from the primary vertex was 0.5 cm.
Minimum bias events with two of more K$^0_s$'s were selected for use
in the analysis, with 19\% having greater than two K$^0_s$'s.
A cut was imposed to prevent K$^0_s$K$^0_s$ pairs from sharing the
same decay daughter.

\begin{figure}
\begin{center}
\includegraphics[width=100mm]{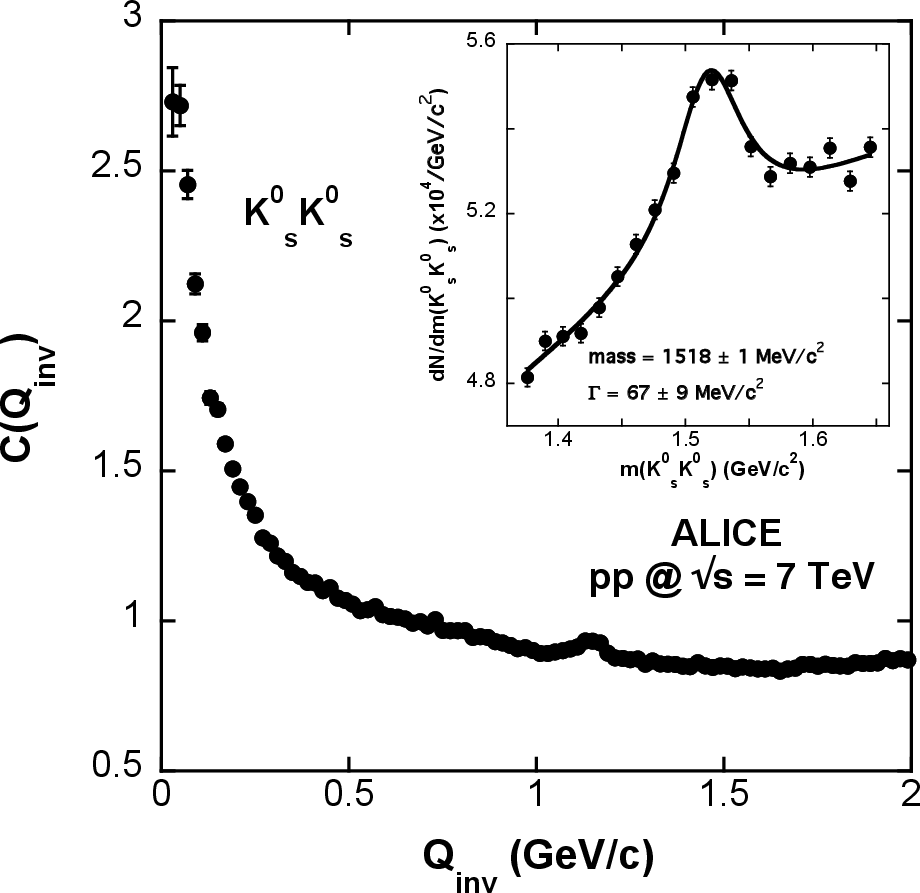} \caption{Inclusive (all event multiplicities
and $k_{\rm T}$)
K$^0_s$K$^0_s$ $Q_{\rm inv}$
correlation function. Plotted in the insert to the f${\rm i}$gure is the invariant K$^0_s$K$^0_s$ mass
distribution, d$N$/d$m$(K$^0_s$K$^0_s$), in the vicinity of the small peak at
$Q_{\rm inv}\approx1.15$ GeV/$c$.}
\label{fig2}
\end{center}
\end{figure}

\section{Results}

Figure \ref{fig2} shows a K$^0_s$K$^0_s$ correlation function in the
invariant momentum difference variable   $Q_{\rm inv}=\sqrt{Q^2-Q_0^2}$,
where $Q$ and $Q_0$ are the 3--momentum and energy differences between
the two particles respectively, for all event multiplicities and $k_{\rm T}$.
The correlation function was formed from the ratio of ``real'' K$^0_s$K$^0_s$
pairs from the same event to ``background'' K$^0_s$K$^0_s$ pairs
constructed by event mixing of ten adjacent events.
Bins in $Q_{\rm inv}$ were taken to be 20 MeV/$c$ which is greater than the average
resolution of $Q_{\rm inv}$ resulting from the
experimental momentum resolution. Also, the enhancement region in $Q_{inv}$
of the correlation functions for source sizes 
of $\sim1$ fm is $\sim200$ MeV/$c$. Thus
the smearing of the correlation function by the experimental momentum resolution
has a negligible effect on the present measurements.
The
three main features seen in this correlation function are 1) a
well-def${\rm i}$ned enhancement region for $Q_{\rm inv}<0.3$ GeV/$c$, 2) a non-f${\rm l}$at
baseline for $Q_{\rm inv}>0.3$ GeV/$c$  and 3) a small peak at
$Q_{\rm inv}\approx1.15$ GeV/$c$.  

Considering feature 3) f${\rm i}$rst, f${\rm i}$tting a
quadratic + Breit-Wigner function to the invariant K$^0_s$K$^0_s$ mass
distribution, d$N$/d$m$(K$^0_s$K$^0_s$), around this peak, where
$m$(K$^0_s$K$^0_s$)$=2\sqrt{(Q_{\rm inv}/2)^2+m_{\rm K}^2}$, we obtain a mass of
$1518 \pm 1 \pm 20$ MeV/$c^2$ and full  width ($\Gamma$) of $67\pm9
\pm10$ MeV/$c^2$ (giving the statistical and systematic errors,
respectively). This is plotted in the insert to Figure \ref{fig2}. Comparing 
with the Particle Data Group meson table
\cite{pdg}, this  peak is a good candidate for the f$_2'$(1525)
meson ($m=1525\pm5$ MeV/$c^2$, $\Gamma=73^{+6}_{-5}$ MeV/$c^2$).
This is the f${\rm i}$rst observation of the decay of this meson
into the K$^0_s$K$^0_s$ channel in pp collisions. 

\begin{figure}
\begin{center}
\includegraphics[width=80mm]{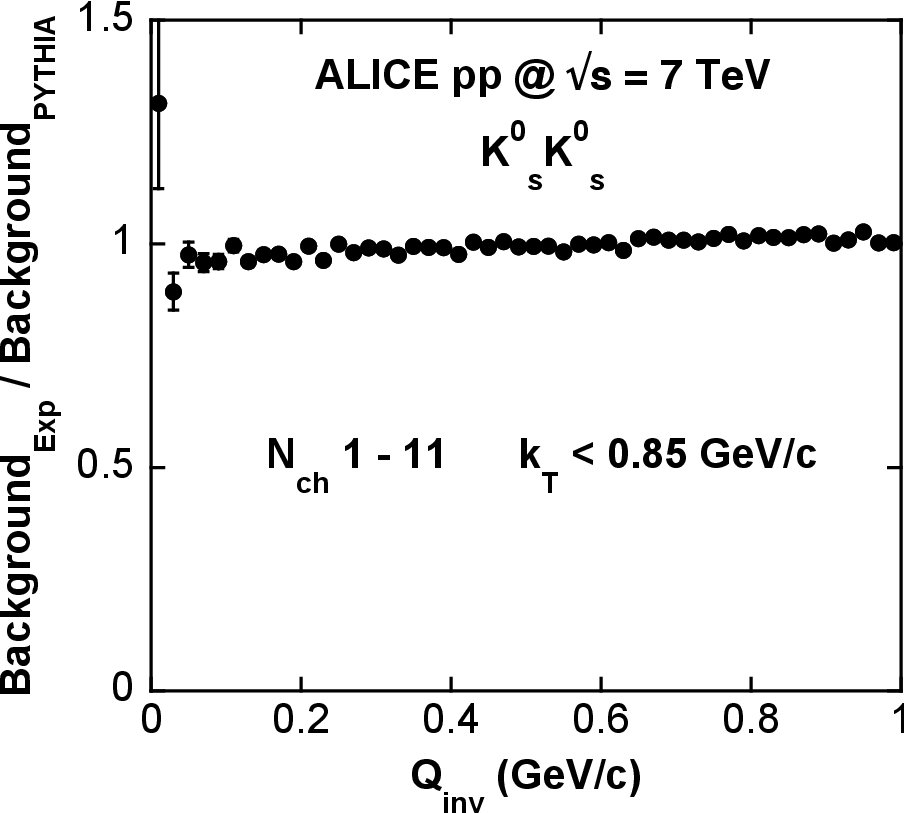} \caption{Ratio of 
K$^0_s$K$^0_s$ experimental background to PYTHIA background vs. $Q_{\rm inv}$
for the range $N_{\rm ch}$ 1-11,
$k_{\rm T}<0.85$ GeV/c.
The average of the ratio is normalized to unity.}
\label{figx}
\end{center}
\end{figure}

In order to
disentangle the non-f${\rm l}$at baseline from the low-$Q_{\rm inv}$ femtoscopic
enhancement, the Monte Carlo event generator PYTHIA
\cite{pythia6.4,perugia} was used to model the baseline. PYTHIA contains
neither quantum statistics nor the K$^0_s$K$^0_s \rightarrow$ a$_0$/f$_0$
channel, but does contain other kinematic effects which could lead to
baseline correlations such as mini-jets and momentum and energy conservation
effects \cite{Aamodt:2011kd}.
PYTHIA events were reconstructed and run through the same analysis method
as used for the corresponding experimental data runs to simulate the
same conditions as the experimental data analysis.
The PYTHIA version of the invariant mass distributions shown for experiment
in Figure \ref{fig1} yielded similar $S/(S+B)$ values.
As a test, the K$^0_s$K$^0_s$ background obtained from event mixing
using PYTHIA events was compared with that from experiment. Since the
background pairs do not have femtoscopic effects, these
should ideally be in close agreement.
A sample plot of the experimental
to PYTHIA ratio of the background vs. $Q_{\rm inv}$ is
shown in Figure \ref{figx} for the range $N_{\rm ch}$ 1-11,
$k_{\rm T}<0.85$ GeV/c. 
The average of the ratio is normalized to unity. It is found that PYTHIA
agrees with the experimental backgrounds within 10\%.
The Monte Carlo event generator PHOJET \cite{Engel:1994vs,Engel:1995yda}
was also studied for use in modeling the baseline, but it was found to not agree
as well with experiment as PYTHIA, and thus was not used.

K$^0_s$K$^0_s$ correlation functions in $Q_{\rm inv}$ were formed from the
data in four ranges: two event multiplicity (1-11,  $>11$) ranges
times two $k_{\rm T}$ ($<0.85$, $>0.85$ GeV/$c$)
ranges. Event multiplicity was def${\rm i}$ned as the number of charged
particles falling into the pseudorapidty range $\left | \eta \right |<0.8$ and
transverse momentum range $0.12<p_{\rm T}<10$ GeV/$c$.
The two event multiplicity ranges used, 1-11 and $>11$, correspond to 
mean charged particle densities,
$\langle$d$N_{\rm ch}$/d$\eta \rangle$,
of 2.8 and 11.1, respectively, with uncertainties of $\sim10\%$.
PYTHIA events were used to estimate $\langle$d$N_{\rm ch}$/d$\eta \rangle$ from the 
mean charged--particle multiplicity in each range as was done
for Table I of Reference \cite{Aamodt:2011kd},
which presents ALICE $\pi\pi$ results for pp collisions at $\sqrt{s}=7$ TeV,
event multiplicity having been determined in the same way there as in the present work.
This is convenient since the K$^0_s$K$^0_s$ source parameters from the 
present work are compared
with those from the $\pi\pi$ measurement below.
About $3\times10^8$
minimum bias events were analyzed yielding about $6\times10^6$
K$^0_s$K$^0_s$ pairs. 
A similar number of PYTHIA minimum bias events used for the baseline
determination were also analyzed. This was found to give suff${\rm i}$cient statistics
for the PYTHIA correlation functions such that the impact of these
statistical uncertainties on the measurement of the source parameters was small compared with
the systematic uncertainties present in the measurement.
The femtoscopic variables $R$ and $\lambda$ were
extracted in each range by f${\rm i}$tting the experimental correlation
function divided by the PYTHIA correlation function with the
Lednicky parametrization \cite{Abelev:2006gu} based on the model by
R. Lednicky and V.L. Lyuboshitz \cite{lednicky}. This model takes into
account both quantum statistics and strong f${\rm i}$nal-state interactions
from the a$_0$/f$_0$ resonance which occur between the K$^0_s$K$^0_s$
pair. The K$^0_s$  spacial distribution is assumed to be Gaussian with
a width $R$ in the parametrization and so its inf${\rm l}$uence on the
correlation function is from both the quantum statistics and the
strong f${\rm i}$nal-state interaction. This is the same parametrization as
was used by the RHIC STAR collaboration to extract $R$ and $\lambda$
from their K$^0_s$K$^0_s$ study  of Au--Au collisions
\cite{Abelev:2006gu}. 
The correlation function is
\begin{eqnarray}
C(Q_{\rm inv})=\lambda C'(Q_{\rm inv})+(1-\lambda)
\end{eqnarray}
where,
\begin{eqnarray}
C'(Q_{\rm inv})=1+e^{-Q_{\rm inv}^2R^2}+\alpha \left [ \left | \frac{f(k^*)}{R} \right |^2 + \frac{4\Re f(k^*)}{\sqrt{\pi }R}F_1(Q_{\rm inv}R)-\frac{2\Im f(k^*)}{R}F_2(Q_{\rm inv}R) \right ]
\end{eqnarray}
and where
\begin{eqnarray}
F_1(z)=\int_{0}^{z}dx\frac{e^{x^2-z^2}}{z} \ ;  F_2(z)=\frac{1-e^{-z^2}}{z}.
\end{eqnarray}
$f(k^*)$ is the s--wave K$^0$$\overline{\rm K^0}$ scattering amplitude whose main
contributions are the
s--wave isoscalar and isovector f$_0$ and a$_0$ resonances \cite{Abelev:2006gu},
$R$ is the radius parameter and $\lambda$ is the correlation strength parameter (in the ideal
case of pure quantum statistics $\lambda=1$). 
$\alpha$ is the fraction of K$^0_s$K$^0_s$ pairs that come from 
the K$^0$$\overline{\rm K^0}$ system which is
set to 0.5 assuming symmetry in K$^0$ and $\overline{\rm K^0}$ production \cite{Abelev:2006gu}.
As seen in Eq. (2), the f${\rm i}$rst term is the usual Gaussian from quantum statistics and the
second term describes the f${\rm i}$nal-state resonance scattering and both are sensitive to
the radius parameter, $R$, giving enhanced sensitivity to this parameter.
The scattering amplitude, $f(k^*)$, depends on the resonance masses and decay
couplings which have been extracted in various experiments \cite{Abelev:2006gu}. 
The uncertainties in these are found to have only a small effect on the extraction of
$R$ and $\lambda$ in the present study. An overall normalization parameter multiplying
Eq. (1) is also f${\rm i}$t to the experimental correlation function.

\begin{figure}
\begin{center}
\includegraphics[width=100mm]{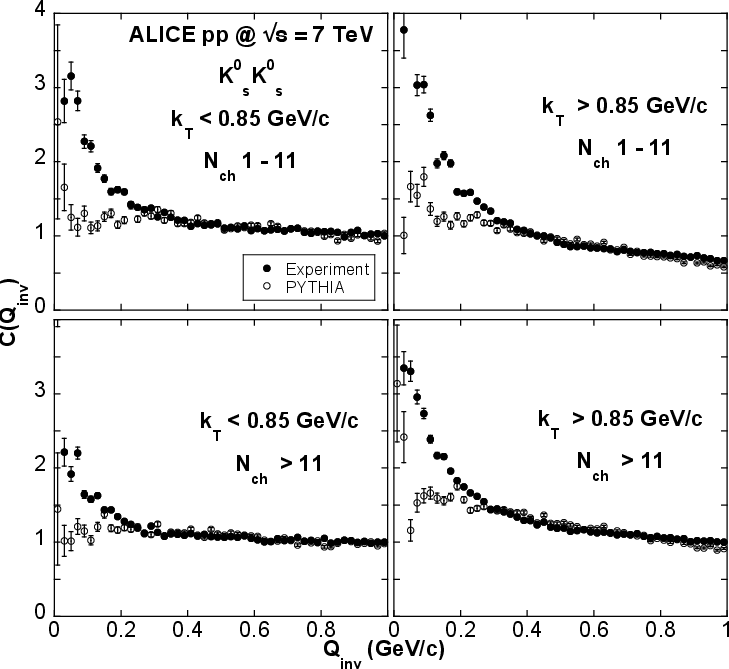} \caption{Experimental 
and PYTHIA K$^0_s$K$^0_s$ 
correlation functions for the
four multiplicity--$k_{\rm T}$ ranges.}
\label{fig3}
\end{center}
\end{figure}

\begin{figure}
\begin{center}
\includegraphics[width=100mm]{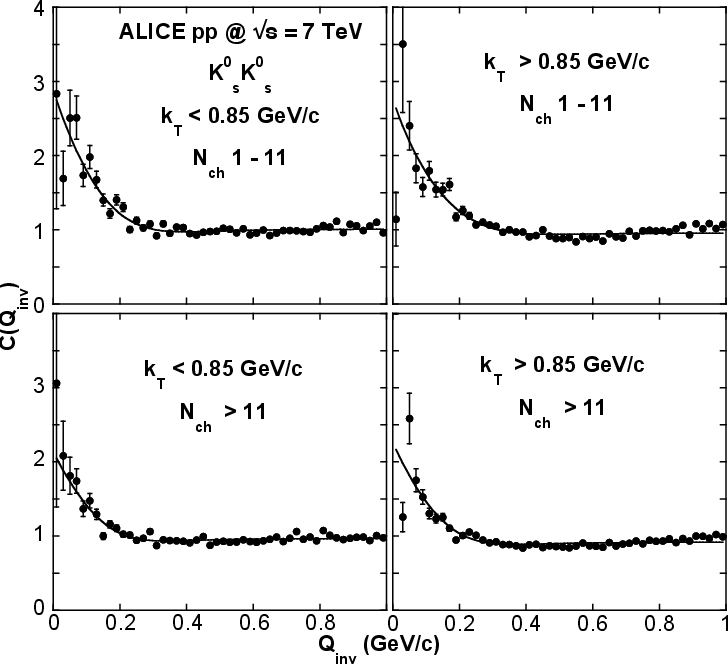} \caption{Experimental K$^0_s$K$^0_s$ 
correlation functions divided by PYTHIA correlation functions for the
four multiplicity--$k_{\rm T}$ ranges
with femtoscopic f${\rm i}$ts using the Lednicky parametrization.}
\label{fig5}
\end{center}
\end{figure}

Figure \ref{fig3} shows the experimental and PYTHIA K$^0_s$K$^0_s$ correlation functions 
for each of the four multiplicity--$k_{\rm T}$ ranges used. Whereas the experimental
correlation functions show an enhancement for $Q_{\rm inv}<0.3$ GeV/$c^2$, the
PYTHIA correlation functions do not show a similar enhancement. This is what would be
expected if the experimental correlation functions contain femtoscopic correlations
since PYTHIA does not contain these. PYTHIA is seen to describe the experimental
baseline rather well in the region $Q_{\rm inv}>0.4$ GeV/$c^2$ where it is
expected that effects of femtoscopic correlations are insignif${\rm i}$cant.  
Figure \ref{fig5} shows the experimental
correlation functions divided by the PYTHIA
correlation functions from Figure \ref{fig3}
along with the f${\rm i}$ts with the Lednicky parametrization from Eqs. (1)-(3).
The f${\rm i}$ts are seen to qualitatively describe the correlation functions within the
error bars, which are statistical.

\begin{figure}
\begin{center}
\includegraphics[width=80mm]{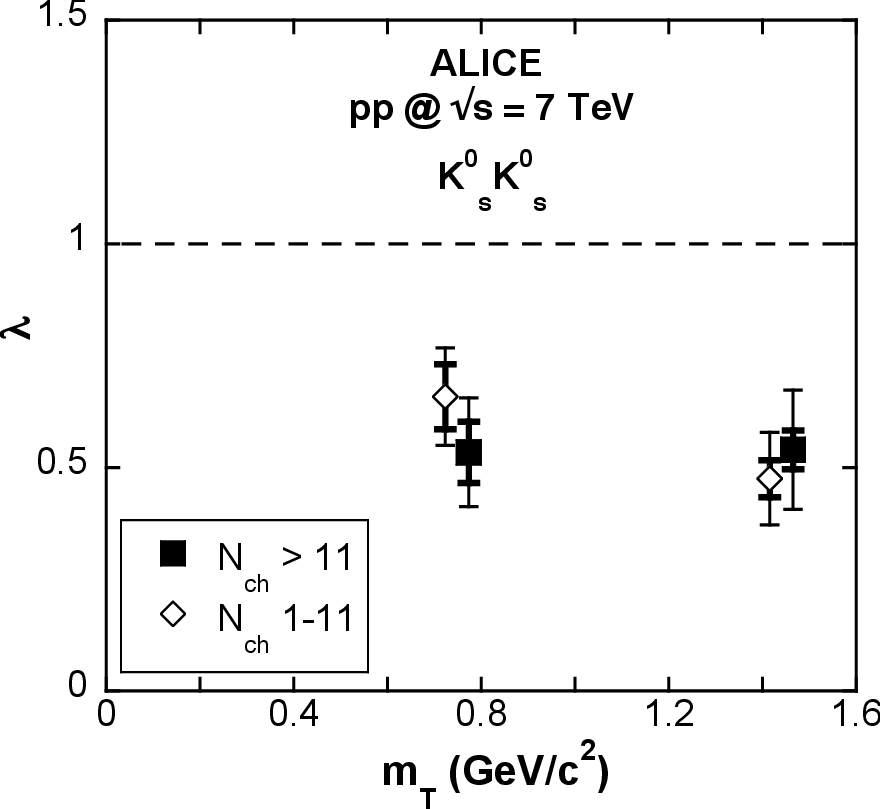} \caption{$\lambda$ parameters
  extracted by f${\rm i}$tting the Lednicky parametrization to
  K$^0_s$K$^0_s$ correlation functions as shown in Figure \ref{fig5}. 
  Statistical (darker lines) and total errors are shown.
  The $N_{\rm ch}>11$ points are offset by 0.05 GeV/$c^2$ for clarity.}
\label{fig6}
\end{center}
\end{figure}

\begin{figure}
\begin{center}
\includegraphics[width=80mm]{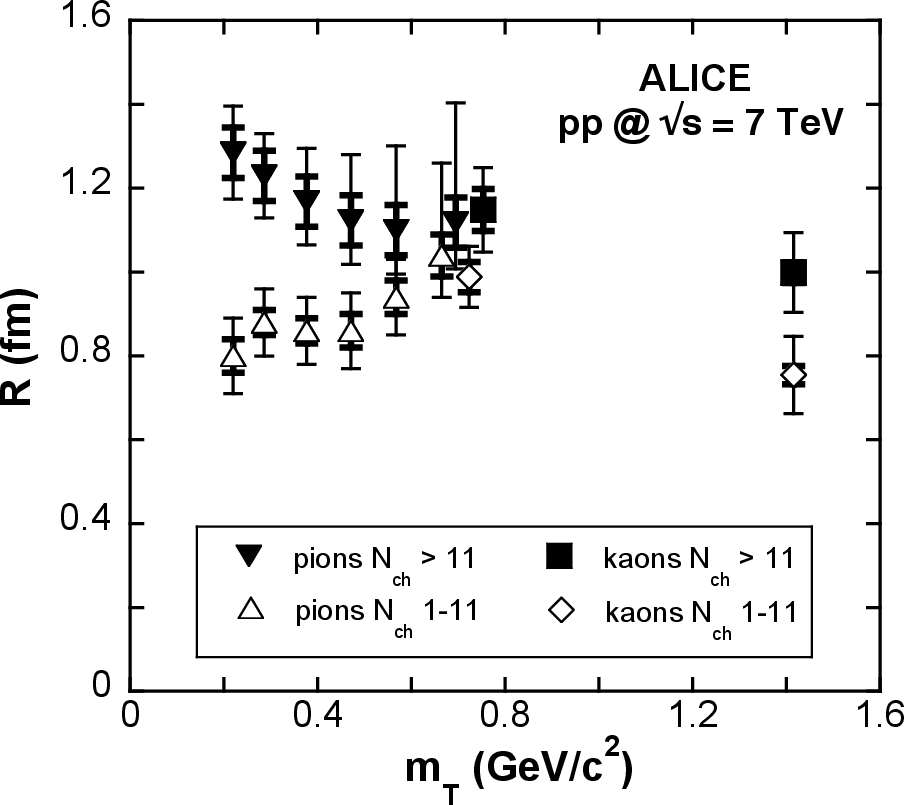} \caption{$R$
  parameters extracted by f${\rm i}$tting the Lednicky parametrization to
  K$^0_s$K$^0_s$ correlation functions as shown in Figure \ref{fig5}. 
  Also shown for comparison are $R$ parameters extracted in
  the same event multiplicity ranges from a $\pi\pi$ femtoscopic study
  by ALICE \cite{Aamodt:2011kd} in pp collisions at $\sqrt{s}=7$ TeV. Statistical (darker lines)
  and total errors are shown.
  The highest $m_{\rm T}$ pion $N_{\rm ch}>11$ point and lower 
  $m_{\rm T}$ kaon $N_{\rm ch}>11$ point have been shifted by 0.03 GeV/$c^2$
  for clarity.} 
\label{fig7}
\end{center}
\end{figure}

Figures \ref{fig6} and \ref{fig7} and Table \ref{table1} present the results of this study
for $\lambda$ and $R$ parameters extracted by f${\rm i}$tting the Lednicky
parametrization to K$^0_s$K$^0_s$ correlation functions as
shown in Figure \ref{fig5}. 
The source parameters are plotted versus 
$m_{\rm T}=\sqrt{\langle k_{\rm T} \rangle^2+m_{\rm K}^2}$
to observe whether $m_{\rm T}$ scaling is present (as discussed earlier)
and statistical + systematic error bars are shown.
The statistical uncertainties include both the experimental and
the PYTHIA statistical uncertainties used to form the correlation functions, as shown in
Figure \ref{fig3}.
The largest
contributions to the systematic uncertainties are 1) the non-statistical
uncertainty in using PYTHIA to determine the baseline and 2) the effect of
varying the $Q_{\rm inv}$ f${\rm i}$t range by $\pm10\%$. These were found to be on the order of
or greater than the size of the statistical uncertainties, as can be seen in
Table \ref{table1}. The method used to estimate the systematic uncertainty
of using PYTHIA was to set the PYTHIA K$^0_s$K$^0_s$ background
distribution equal to the experimental background distribution in the ratio
of correlation functions, e.g.  forcing the ratio plotted in Figure \ref{figx} to be
exactly unity for all $Q_{\rm inv}$.
The ratio of correlation functions then becomes
the ratio of
the experimental to PYTHIA real pair distributions, which is then f${\rm i}$t 
with the Lednicky parametrization to extract the source parameters.
Parameters extracted from these correlation functions were then 
averaged with those from Figure \ref{fig5} and are given in
Figures \ref{fig6} and \ref{fig7} and Table \ref{table1}.
This method is similar to that used in estimating systematic uncertainties in
other K$^0_s$K$^0_s$ measurements \cite{Schael:2004qn, Chekanov:2007ev}.

To see the effect of the $a_0/f_0$ f${\rm i}$nal-state interaction (FSI) term in the 
Lednicky parametrization, the correlation
functions in Figure \ref{fig5} were f${\rm i}$t with Eqs. (1)-(3) for two cases: 1) quantum statistics +
FSI terms, i.e. $\alpha=0.5$ in Eq. (2), and 2) quantum statistics
term only, i.e. $\alpha=0$ in Eq. (2). Case 2) corresponds to the usual Gaussian
parametrization for $R$ and $\lambda$. The results of these f${\rm i}$ts are shown in
Table \ref{table2}.
Including the FSI term in the f${\rm i}$t is seen to signif${\rm i}$cantly reduce both $R$ and $\lambda$,
i.e. $R$ by $\sim30\%$ and $\lambda$ by $\sim50\%$.
The FSI is thus seen to enhance the correlation function for $Q_{\rm inv}\rightarrow 0$
making $\lambda$ appear larger and making the enhancement region narrower
resulting in an apparent larger $R$. 
A reduction in $R$ and $\lambda$ when including the FSI term was
also observed, but to a lesser extent, in the STAR Au--Au K$^0_s$K$^0_s$ 
study \cite{Abelev:2006gu}. A larger effect of the a$_0$/f$_0$ resonance 
on the correlation function in pp collisions compared with Au--Au collisions is
expected since the
two kaons are produced in closer proximity to each other in pp collisions, enhancing
the probability for f${\rm i}$nal-state interactions.

Within the uncertainties, the $m_{\rm T}$
dependence of $\lambda$ is seen in Figure \ref{fig6} to be mostly f${\rm l}$at
with $\lambda$ lying at an average level of $\sim0.5-0.6$, similar to
that found in  the ALICE $\pi\pi$ results for pp collisions at $\sqrt{s}=7$ TeV
\cite{Aamodt:2011kd}.  In $\pi\pi$ studies the $\lambda$ smaller than 1 has been 
shown at least in part to be due to the presence of long--lived meson resonances
which distort the shape of the source so that the Gaussian assumption,
which the f${\rm i}$tting functions are based on, is less valid
\cite{Humanic:2006ib}. This same explanation is possible for the
present $\lambda$ parameters extracted from the K$^0_s$K$^0_s$ correlation functions.
For example, the
$\phi$ and K$^*$ mesons with full widths of 
$\Gamma\sim4$ and $\Gamma\sim50$ MeV/$c^2$, respectively,
could act as
long--lived resonances compared with the extracted source scale of
$R\sim1$ fm, the larger scales being unresolved in the
f${\rm i}$rst few $Q_{\rm inv}$ bins but still depressing the overall correlation function.

\begin{table*}
\caption{K$^0_s$K$^0_s$ source parameters from Lednicky f${\rm i}$ts
for $\sqrt{s}=7$ TeV pp collisions. Statistical and systematic errors are shown.\label{table1}}
\begin{tabular}{cccccc}
$k_{\rm T}$ range&$N_{\rm ch}$ range&$\langle k_{\rm T} \rangle$&$\langle$d$N_{\rm ch}$/d$\eta \rangle$&$\lambda$&$R$ \\
(GeV/$c$)&   &(GeV/$c$)&  &   &(fm) \\ \hline
$<0.85$&$1-11$&$0.52$&$2.8$&$0.66\pm0.07\pm0.04$&$0.99\pm0.04\pm0.04$ \\
$>0.85$&$1-11$&$1.32$&$2.8$&$0.48\pm0.04\pm0.06$&$0.75\pm0.02\pm0.07$ \\
$<0.85$&$>11$&$0.52$&$11.1$&$0.53\pm0.07\pm0.05$&$1.15\pm0.05\pm0.05$ \\
$>0.85$&$>11$&$1.32$&$11.1$&$0.54\pm0.04\pm0.09$&$1.00\pm0.02\pm0.07$ \\
\end{tabular}
\end{table*}

\begin{table*}
\caption{K$^0_s$K$^0_s$ source parameters comparing $\alpha=0.5$
(quantum statistics+FSI) and $\alpha=0$ (quantum statistics only) f${\rm i}$ts
to Figure \ref{fig5} using Eqs. (1)-(3). Statistical errors are shown.\label{table2}}
\begin{tabular}{cccccc}
$k_{\rm T}$ range&$N_{\rm ch}$ range&$\lambda$&$R$&$\lambda$&$R$ \\
(GeV/$c$)&   &   &(fm)&   &(fm) \\ 
  &  &$\alpha=0.5$&$\alpha=0.5$&$\alpha=0$&$\alpha=0$ \\ \hline
$<0.85$&$1-11$&$0.64\pm0.07$&$0.96\pm0.04$&$1.36\pm0.15$&$1.35\pm0.07$ \\
$>0.85$&$1-11$&$0.50\pm0.04$&$0.81\pm0.02$&$1.07\pm0.09$&$1.05\pm0.04$ \\
$<0.85$&$>11$&$0.51\pm0.07$&$1.12\pm0.05$&$0.97\pm0.15$&$1.64\pm0.11$ \\
$>0.85$&$>11$&$0.56\pm0.05$&$1.03\pm0.02$&$0.89\pm0.10$&$1.37\pm0.07$ \\
\end{tabular}
\end{table*}

In Figure \ref{fig7} the dependence of 
the extracted radius parameters on the transverse mass and event multiplicity are shown.
Also shown for comparison are
$R$ parameters extracted in the same event multiplicity ranges from a
$\pi\pi$ femtoscopic study by  ALICE \cite{Aamodt:2011kd} in 7 TeV
pp collisions. 
Looking at the $m_{\rm T}$ dependence f${\rm i}$rst,
the K$^0_s$K$^0_s$ results alone suggest a
tendency for $R$ to decrease with increasing $m_{\rm T}$
for both multiplicity ranges. The $\pi\pi$ measurements also show this
decreasing trend for the high multiplicity range, but show the opposite
trend for the low multiplicity range, $R$ increasing slightly for increasing
$m_{\rm T}$.
Taken with the $\pi\pi$ results the K$^0_s$K$^0_s$
results for $R$ extend the covered range of $m_{\rm T}$ to $\sim1.3$ GeV/$c$,
which is more than twice the range as for $\pi\pi$.
The lower $m_{\rm T}$ points for K$^0_s$K$^0_s$ which are in close 
proximity in $m_{\rm T}$ to the highest $m_{\rm T}$
points for $\pi\pi$ are seen to overlap within errors, showing $m_{\rm T}$ scaling.
The  $m_{\rm T}$ dependence of $R$ combining both particle species is seen 
to be weak or nonexistent within the error bars.
Looking at the multiplicity dependence, a
tendency for $R$ to increase overall for
increasing event multiplicity is seen for both $\pi\pi$ and  K$^0_s$K$^0_s$ measurements
as is observed in $\pi\pi$
heavy-ion collision studies
\cite{Aamodt:2011mr}.

The multiplicity--$m_{\rm T}$ dependence of the pion femtoscopic radii in
heavy--ion collisions is interpreted as a signature for collective
hydrodynamic matter behaviour~\cite{Lisa:2005a}. 
The corresponding measurements
in pp collisions at $\sqrt{s}=7$~TeV show similar
behaviour~\cite{Aamodt:2011kd}.
However, important differences with heavy-ion collisions remain,
for example the low--multiplicity $m_{\rm T}$ dependence of $R$ in pp for
pions seems to increase with increasing $m_{\rm T}$ rather than decreasing
as with heavy--ion collisions,
as already mentioned earlier.
The interpretation of these pp results for pions is
still not clear, although model calculations exist that attempt to
explain them via a collective phase created in high--multiplicity pp
collisions~\cite{Bozek:2009dt,Werner:2011fd,truesdale}. 
If such a collective phase
is hydrodynamic--like, the $m_{\rm T}$ dependence of the radii should extend to
heavier particles such as the K$^0_s$ as well, as shown 
in Ref. \cite{truesdale}. The measurements
presented in this paper provide a crucial cross-check of the
collectivity hypothesis. The interpretation is, however, complicated by
the fact that in such small systems particles coming from the decay of
strong resonances play a signif${\rm i}$cant role~\cite{Kisiel:2010xy}; simple
chemical model calculations show that this inf${\rm l}$uence should be
relatively smaller for kaons than for pions.
So far, no model calculations are known
in the literature for any KK correlations in pp
collisions for $m_{\rm T}\geq 0.7$ GeV/$c^2$, but the results measured 
in the present study should act as a motivation for such calculations.

\section{Summary}

In summary, identical neutral kaon pair correlations have been measured 
in $\sqrt{s}=7$
TeV pp  collisions in the ALICE experiment. One--dimensional
K$^0_s$K$^0_s$ correlation functions in terms of the invariant
momentum difference of kaon pairs were formed in two multiplicity
and two transverse momentum ranges. The femtoscopic kaon source
parameters $R$ and $\lambda$ have been extracted. The f${\rm i}$t includes quantum
statistics and f${\rm i}$nal--state interactions of the a$_0$/f$_0$
resonance. K$^0_s$K$^0_s$ correlations show an increase in $R$
for increasing multiplicity and a slight decrease in $R$ for
increasing $m_{\rm T}$ as seen in $\pi\pi$ correlations in the pp system
and in heavy--ion collisions.
The universality of the $m_{\rm T}$ dependence of the extracted radii,
i.e. $m_{\rm T}$ scaling, is also observed within uncertainties for
the K$^0_s$K$^0_s$ and $\pi\pi$ radii.

\newenvironment{acknowledgement}{\relax}{\relax}
\begin{acknowledgement}
\section{Acknowledgements}
The ALICE collaboration would like to thank all its engineers and technicians for their invaluable contributions to the construction of the experiment and the CERN accelerator teams for the outstanding performance of the LHC complex.
\\
The ALICE collaboration acknowledges the following funding agencies for their support in building and
running the ALICE detector:
 \\
Calouste Gulbenkian Foundation from Lisbon and Swiss Fonds Kidagan, Armenia;
 \\
Conselho Nacional de Desenvolvimento Cient\'{\i}f${\rm i}$co e Tecnol\'{o}gico (CNPq), Financiadora de Estudos e Projetos (FINEP),
Funda\c{c}\~{a}o de Amparo \`{a} Pesquisa do Estado de S\~{a}o Paulo (FAPESP);
 \\
National Natural Science Foundation of China (NSFC), the Chinese Ministry of Education (CMOE)
and the Ministry of Science and Technology of China (MSTC);
 \\
Ministry of Education and Youth of the Czech Republic;
 \\
Danish Natural Science Research Council, the Carlsberg Foundation and the Danish National Research Foundation;
 \\
The European Research Council under the European Community's Seventh Framework Programme;
 \\
Helsinki Institute of Physics and the Academy of Finland;
 \\
French CNRS-IN2P3, the `Region Pays de Loire', `Region Alsace', `Region Auvergne' and CEA, France;
 \\
German BMBF and the Helmholtz Association;
\\
General Secretariat for Research and Technology, Ministry of
Development, Greece;
\\
Hungarian OTKA and National Off${\rm i}$ce for Research and Technology (NKTH);
 \\
Department of Atomic Energy and Department of Science and Technology of the Government of India;
 \\
Istituto Nazionale di Fisica Nucleare (INFN) of Italy;
 \\
MEXT Grant-in-Aid for Specially Promoted Research, Ja\-pan;
 \\
Joint Institute for Nuclear Research, Dubna;
 \\
National Research Foundation of Korea (NRF);
 \\
CONACYT, DGAPA, M\'{e}xico, ALFA-EC and the HELEN Program (High-Energy physics Latin-American--European Network);
 \\
Stichting voor Fundamenteel Onderzoek der Materie (FOM) and the Nederlandse Organisatie voor Wetenschappelijk Onderzoek (NWO), Netherlands;
 \\
Research Council of Norway (NFR);
 \\
Polish Ministry of Science and Higher Education;
 \\
National Authority for Scientif${\rm i}$c Research - NASR (Autoritatea Na\c{t}ional\u{a} pentru Cercetare \c{S}tiin\c{t}if${\rm i}$c\u{a} - ANCS);
 \\
Federal Agency of Science of the Ministry of Education and Science of Russian Federation, International Science and
Technology Center, Russian Academy of Sciences, Russian Federal Agency of Atomic Energy, Russian Federal Agency for Science and Innovations and CERN-INTAS;
 \\
Ministry of Education of Slovakia;
 \\
Department of Science and Technology, South Africa;
 \\
CIEMAT, EELA, Ministerio de Educaci\'{o}n y Ciencia of Spain, Xunta de Galicia (Conseller\'{\i}a de Educaci\'{o}n),
CEA\-DEN, Cubaenerg\'{\i}a, Cuba, and IAEA (International Atomic Energy Agency);
 \\
Swedish Research Council (VR) and Knut $\&$ Alice Wallenberg
Foundation (KAW);
 \\
Ukraine Ministry of Education and Science;
 \\
United Kingdom Science and Technology Facilities Council (STFC);
 \\
The United States Department of Energy, the United States National
Science Foundation, the State of Texas, and the State of Ohio.

\end{acknowledgement}
\newpage
%
%
\appendix
\section{The ALICE Collaboration}
\label{app:collab}

\begingroup
\small
\begin{flushleft}
B.~Abelev\Irefn{org1234}\And
J.~Adam\Irefn{org1274}\And
D.~Adamov\'{a}\Irefn{org1283}\And
A.M.~Adare\Irefn{org1260}\And
M.M.~Aggarwal\Irefn{org1157}\And
G.~Aglieri~Rinella\Irefn{org1192}\And
A.G.~Agocs\Irefn{org1143}\And
A.~Agostinelli\Irefn{org1132}\And
S.~Aguilar~Salazar\Irefn{org1247}\And
Z.~Ahammed\Irefn{org1225}
% ...
\renewcommand\labelenumi{\textsuperscript{\theenumi}~}
\section*{Aff${\rm i}$liation notes}
\renewcommand\theenumi{\roman{enumi}}
\begin{Authlist}
\item \Adef{M.V.Lomonosov Moscow State University, D.V.Skobeltsyn Institute of Nuclear Physics, Moscow, Russia}Also at: M.V.Lomonosov Moscow State University, D.V.Skobeltsyn Institute of Nuclear Physics, Moscow, Russia
\end{Authlist}
\section*{Collaboration Institutes}
\renewcommand\theenumi{\arabic{enumi}~}
\begin{Authlist}
\item \Idef{org1279}Benem\'{e}rita Universidad Aut\'{o}noma de Puebla, Puebla, Mexico
\item \Idef{org1220}Bogolyubov Institute for Theoretical Physics, Kiev, Ukraine
\item \Idef{org1262}Budker Institute for Nuclear Physics, Novosibirsk, Russia
\item \Idef{org1292}California Polytechnic State University, San Luis Obispo, California, United States
\item \Idef{org14939}Centre de Calcul de l'IN2P3, Villeurbanne, France
\item \Idef{org1197}Centro de Aplicaciones Tecnol\'{o}gicas y Desarrollo Nuclear (CEADEN), Havana, Cuba
\item \Idef{org1242}Centro de Investigaciones Energ\'{e}ticas Medioambientales y Tecnol\'{o}gicas (CIEMAT), Madrid, Spain
\item \Idef{org1244}Centro de Investigaci\'{o}n y de Estudios Avanzados (CINVESTAV), Mexico City and M\'{e}rida, Mexico
\item \Idef{org1335}Centro Fermi -- Centro Studi e Ricerche e Museo Storico della Fisica ``Enrico Fermi'', Rome, Italy
\item \Idef{org17347}Chicago State University, Chicago, United States
\item \Idef{org1288}Commissariat \`{a} l'Energie Atomique, IRFU, Saclay, France
\item \Idef{org1294}Departamento de F\'{\i}sica de Part\'{\i}culas and IGFAE, Universidad de Santiago de Compostela, Santiago de Compostela, Spain
\item \Idef{org1106}Department of Physics Aligarh Muslim University, Aligarh, India
\item \Idef{org1121}Department of Physics and Technology, University of Bergen, Bergen, Norway
\item \Idef{org1162}Department of Physics, Ohio State University, Columbus, Ohio, United States
\item \Idef{org1300}Department of Physics, Sejong University, Seoul, South Korea
\item \Idef{org1268}Department of Physics, University of Oslo, Oslo, Norway
\item \Idef{org1145}Dipartimento di Fisica dell'Universit\`{a} and Sezione INFN, Cagliari, Italy
\item \Idef{org1270}Dipartimento di Fisica dell'Universit\`{a} and Sezione INFN, Padova, Italy
\item \Idef{org1315}Dipartimento di Fisica dell'Universit\`{a} and Sezione INFN, Trieste, Italy
\item \Idef{org1132}Dipartimento di Fisica dell'Universit\`{a} and Sezione INFN, Bologna, Italy
\item \Idef{org1285}Dipartimento di Fisica dell'Universit\`{a} `La Sapienza' and Sezione INFN, Rome, Italy
\item \Idef{org1154}Dipartimento di Fisica e Astronomia dell'Universit\`{a} and Sezione INFN, Catania, Italy
\item \Idef{org1290}Dipartimento di Fisica `E.R.~Caianiello' dell'Universit\`{a} and Gruppo Collegato INFN, Salerno, Italy
\item \Idef{org1312}Dipartimento di Fisica Sperimentale dell'Universit\`{a} and Sezione INFN, Turin, Italy
\item \Idef{org1103}Dipartimento di Scienze e Innovazione Tecnologica dell'Universit\`{a} del Piemonte Orientale and Gruppo Collegato INFN, Alessandria, Italy
\item \Idef{org1114}Dipartimento Interateneo di Fisica `M.~Merlin' and Sezione INFN, Bari, Italy
\item \Idef{org1237}Division of Experimental High Energy Physics, University of Lund, Lund, Sweden
\item \Idef{org1192}European Organization for Nuclear Research (CERN), Geneva, Switzerland
\item \Idef{org1227}Fachhochschule K\"{o}ln, K\"{o}ln, Germany
\item \Idef{org1122}Faculty of Engineering, Bergen University College, Bergen, Norway
\item \Idef{org1136}Faculty of Mathematics, Physics and Informatics, Comenius University, Bratislava, Slovakia
\item \Idef{org1274}Faculty of Nuclear Sciences and Physical Engineering, Czech Technical University in Prague, Prague, Czech Republic
\item \Idef{org1229}Faculty of Science, P.J.~\v{S}af\'{a}rik University, Ko\v{s}ice, Slovakia
\item \Idef{org1184}Frankfurt Institute for Advanced Studies, Johann Wolfgang Goethe-Universit\"{a}t Frankfurt, Frankfurt, Germany
\item \Idef{org1215}Gangneung-Wonju National University, Gangneung, South Korea
\item \Idef{org1212}Helsinki Institute of Physics (HIP) and University of Jyv\"{a}skyl\"{a}, Jyv\"{a}skyl\"{a}, Finland
\item \Idef{org1203}Hiroshima University, Hiroshima, Japan
\item \Idef{org1329}Hua-Zhong Normal University, Wuhan, China
\item \Idef{org1254}Indian Institute of Technology, Mumbai, India
\item \Idef{org36378}Indian Institute of Technology Indore (IIT), Indore, India
\item \Idef{org1266}Institut de Physique Nucl\'{e}aire d'Orsay (IPNO), Universit\'{e} Paris-Sud, CNRS-IN2P3, Orsay, France
\item \Idef{org1277}Institute for High Energy Physics, Protvino, Russia
\item \Idef{org1249}Institute for Nuclear Research, Academy of Sciences, Moscow, Russia
\item \Idef{org1320}Nikhef, National Institute for Subatomic Physics and Institute for Subatomic Physics of Utrecht University, Utrecht, Netherlands
\item \Idef{org1250}Institute for Theoretical and Experimental Physics, Moscow, Russia
\item \Idef{org1230}Institute of Experimental Physics, Slovak Academy of Sciences, Ko\v{s}ice, Slovakia
\item \Idef{org1127}Institute of Physics, Bhubaneswar, India
\item \Idef{org1275}Institute of Physics, Academy of Sciences of the Czech Republic, Prague, Czech Republic
\item \Idef{org1139}Institute of Space Sciences (ISS), Bucharest, Romania
\item \Idef{org27399}Institut f\"{u}r Informatik, Johann Wolfgang Goethe-Universit\"{a}t Frankfurt, Frankfurt, Germany
\item \Idef{org1185}Institut f\"{u}r Kernphysik, Johann Wolfgang Goethe-Universit\"{a}t Frankfurt, Frankfurt, Germany
\item \Idef{org1177}Institut f\"{u}r Kernphysik, Technische Universit\"{a}t Darmstadt, Darmstadt, Germany
\item \Idef{org1256}Institut f\"{u}r Kernphysik, Westf\"{a}lische Wilhelms-Universit\"{a}t M\"{u}nster, M\"{u}nster, Germany
\item \Idef{org1246}Instituto de Ciencias Nucleares, Universidad Nacional Aut\'{o}noma de M\'{e}xico, Mexico City, Mexico
\item \Idef{org1247}Instituto de F\'{\i}sica, Universidad Nacional Aut\'{o}noma de M\'{e}xico, Mexico City, Mexico
\item \Idef{org23333}Institut of Theoretical Physics, University of Wroclaw
\item \Idef{org1308}Institut Pluridisciplinaire Hubert Curien (IPHC), Universit\'{e} de Strasbourg, CNRS-IN2P3, Strasbourg, France
\item \Idef{org1182}Joint Institute for Nuclear Research (JINR), Dubna, Russia
\item \Idef{org1143}KFKI Research Institute for Particle and Nuclear Physics, Hungarian Academy of Sciences, Budapest, Hungary
\item \Idef{org1199}Kirchhoff-Institut f\"{u}r Physik, Ruprecht-Karls-Universit\"{a}t Heidelberg, Heidelberg, Germany
\item \Idef{org20954}Korea Institute of Science and Technology Information, Daejeon, South Korea
\item \Idef{org1160}Laboratoire de Physique Corpusculaire (LPC), Clermont Universit\'{e}, Universit\'{e} Blaise Pascal, CNRS--IN2P3, Clermont-Ferrand, France
\item \Idef{org1194}Laboratoire de Physique Subatomique et de Cosmologie (LPSC), Universit\'{e} Joseph Fourier, CNRS-IN2P3, Institut Polytechnique de Grenoble, Grenoble, France
\item \Idef{org1187}Laboratori Nazionali di Frascati, INFN, Frascati, Italy
\item \Idef{org1232}Laboratori Nazionali di Legnaro, INFN, Legnaro, Italy
\item \Idef{org1125}Lawrence Berkeley National Laboratory, Berkeley, California, United States
\item \Idef{org1234}Lawrence Livermore National Laboratory, Livermore, California, United States
\item \Idef{org1251}Moscow Engineering Physics Institute, Moscow, Russia
\item \Idef{org1140}National Institute for Physics and Nuclear Engineering, Bucharest, Romania
\item \Idef{org1165}Niels Bohr Institute, University of Copenhagen, Copenhagen, Denmark
\item \Idef{org1109}Nikhef, National Institute for Subatomic Physics, Amsterdam, Netherlands
\item \Idef{org1283}Nuclear Physics Institute, Academy of Sciences of the Czech Republic, \v{R}e\v{z} u Prahy, Czech Republic
\item \Idef{org1264}Oak Ridge National Laboratory, Oak Ridge, Tennessee, United States
\item \Idef{org1189}Petersburg Nuclear Physics Institute, Gatchina, Russia
\item \Idef{org1170}Physics Department, Creighton University, Omaha, Nebraska, United States
\item \Idef{org1157}Physics Department, Panjab University, Chandigarh, India
\item \Idef{org1112}Physics Department, University of Athens, Athens, Greece
\item \Idef{org1152}Physics Department, University of Cape Town, iThemba LABS, Cape Town, South Africa
\item \Idef{org1209}Physics Department, University of Jammu, Jammu, India
\item \Idef{org1207}Physics Department, University of Rajasthan, Jaipur, India
\item \Idef{org1200}Physikalisches Institut, Ruprecht-Karls-Universit\"{a}t Heidelberg, Heidelberg, Germany
\item \Idef{org1325}Purdue University, West Lafayette, Indiana, United States
\item \Idef{org1281}Pusan National University, Pusan, South Korea
\item \Idef{org1176}Research Division and ExtreMe Matter Institute EMMI, GSI Helmholtzzentrum f\"ur Schwerionenforschung, Darmstadt, Germany
\item \Idef{org1334}Rudjer Bo\v{s}kovi\'{c} Institute, Zagreb, Croatia
\item \Idef{org1298}Russian Federal Nuclear Center (VNIIEF), Sarov, Russia
\item \Idef{org1252}Russian Research Centre Kurchatov Institute, Moscow, Russia
\item \Idef{org1224}Saha Institute of Nuclear Physics, Kolkata, India
\item \Idef{org1130}School of Physics and Astronomy, University of Birmingham, Birmingham, United Kingdom
\item \Idef{org1338}Secci\'{o}n F\'{\i}sica, Departamento de Ciencias, Pontif${\rm i}$cia Universidad Cat\'{o}lica del Per\'{u}, Lima, Peru
\item \Idef{org1316}Sezione INFN, Trieste, Italy
\item \Idef{org1271}Sezione INFN, Padova, Italy
\item \Idef{org1313}Sezione INFN, Turin, Italy
\item \Idef{org1286}Sezione INFN, Rome, Italy
\item \Idef{org1146}Sezione INFN, Cagliari, Italy
\item \Idef{org1133}Sezione INFN, Bologna, Italy
\item \Idef{org1115}Sezione INFN, Bari, Italy
\item \Idef{org1155}Sezione INFN, Catania, Italy
\item \Idef{org1322}Soltan Institute for Nuclear Studies, Warsaw, Poland
\item \Idef{org36377}Nuclear Physics Group, STFC Daresbury Laboratory, Daresbury, United Kingdom
\item \Idef{org1258}SUBATECH, Ecole des Mines de Nantes, Universit\'{e} de Nantes, CNRS-IN2P3, Nantes, France
\item \Idef{org1304}Technical University of Split FESB, Split, Croatia
\item \Idef{org1168}The Henryk Niewodniczanski Institute of Nuclear Physics, Polish Academy of Sciences, Cracow, Poland
\item \Idef{org17361}The University of Texas at Austin, Physics Department, Austin, TX, United States
\item \Idef{org1173}Universidad Aut\'{o}noma de Sinaloa, Culiac\'{a}n, Mexico
\item \Idef{org1296}Universidade de S\~{a}o Paulo (USP), S\~{a}o Paulo, Brazil
\item \Idef{org1149}Universidade Estadual de Campinas (UNICAMP), Campinas, Brazil
\item \Idef{org1239}Universit\'{e} de Lyon, Universit\'{e} Lyon 1, CNRS/IN2P3, IPN-Lyon, Villeurbanne, France
\item \Idef{org1205}University of Houston, Houston, Texas, United States
\item \Idef{org20371}University of Technology and Austrian Academy of Sciences, Vienna, Austria
\item \Idef{org1222}University of Tennessee, Knoxville, Tennessee, United States
\item \Idef{org1310}University of Tokyo, Tokyo, Japan
\item \Idef{org1318}University of Tsukuba, Tsukuba, Japan
\item \Idef{org21360}Eberhard Karls Universit\"{a}t T\"{u}bingen, T\"{u}bingen, Germany
\item \Idef{org1225}Variable Energy Cyclotron Centre, Kolkata, India
\item \Idef{org1306}V.~Fock Institute for Physics, St. Petersburg State University, St. Petersburg, Russia
\item \Idef{org1323}Warsaw University of Technology, Warsaw, Poland
\item \Idef{org1179}Wayne State University, Detroit, Michigan, United States
\item \Idef{org1260}Yale University, New Haven, Connecticut, United States
\item \Idef{org1332}Yerevan Physics Institute, Yerevan, Armenia
\item \Idef{org15649}Yildiz Technical University, Istanbul, Turkey
\item \Idef{org1301}Yonsei University, Seoul, South Korea
\item \Idef{org1327}Zentrum f\"{u}r Technologietransfer und Telekommunikation (ZTT), Fachhochschule Worms, Worms, Germany
\end{Authlist}
\endgroup

\end{document}